\documentclass[letterpaper,twocolumn,prl,superscriptaddress]{revtex4}
\usepackage{epsfig}
\usepackage{times}
\usepackage{amsfonts}
\usepackage{color}

\newcommand{\abs}[1]{\ensuremath{\left\vert#1\right\vert}}\newcommand{\ket}[1]{\ensuremath{\vert#1\rangle}}\newcommand{\bra}[1]{\ensuremath{\langle #1\vert}}\newcommand{\kb}[2]{\ensuremath{\vert #1 \rangle \langle #2 \vert}}\newcommand{\exx}[1]{\ensuremath{\langle #1 \rangle}}

\newcommand{\sz}[0]{\ensuremath{\mathbf{\sigma}_z}}\newcommand{\sx}[0]{\ensuremath{\mathbf{\sigma}_x}}\newcommand{\sm}[0]{\ensuremath{\mathbf{\sigma}_{-}}}

\newcommand{\an}[1]{\ensuremath{\hat{#1}}}\newcommand{\cre}[1]{\ensuremath{\hat{#1}^\dagger}}

\newcommand{\ddt}[0]{\frac{\mathrm{d}}{\mathrm{d}t}}\renewcommand{\vec}[1]{\ensuremath{{\mathbf{#1}}}}

\begin{document}

\title{Heat pumping with optically driven excitons}

\author{Erik M Gauger}
\email{erik.gauger@materials.ox.ac.uk}
\affiliation{Department of Materials, University of Oxford, Oxford OX1 3PH, United Kingdom}

\author{Joachim Wabnig}
\affiliation{Department of Materials, University of Oxford, Oxford OX1 3PH, United Kingdom}
\affiliation{Cavendish Laboratory, University of Cambridge, Cambridge CB3 0HE, United Kingdom}

\begin{abstract}
We present a theoretical study showing that an optically driven excitonic
two-level system in a solid state environment can act as a heat pump
by means of repeated phonon emission or absorption events. We derive
a master equation for the combined phonon bath and two-level system
dynamics and analyze the direction and rate of energy transfer as
a function of the externally accessible driving parameters in the coherent control regime. We discover
that if the driving laser is detuned from the exciton transition,
cooling the phonon environment becomes possible. 
\end{abstract}

\maketitle
In semiconductor quantum dots (QDs) the ground state and the state containing a trapped electron-hole pair (exciton) form a two level system  
(2LS) which is a popular implementation of a qubit. Unlike their atomic counterparts, such excitonic qubits are inextricably coupled to the lattice dynamics of the surrounding material \citep{mahan00, borri01, verzelen02, borri05, brandes05}. Longitudinal optical (LO) phonons give rise to excitonic polarons \cite{verzelen02} and are known to play an important role for ultrafast excitation \cite{krummheuer02, grodecka05}. However, in the much slower coherent control regime that is so interesting for quantum information processing, optical phonons only contribute negligibly to the dephasing, and the coupling to longitudinal acoustic (LA) phonons via the deformation potential is predicted to be dominant at low temperatures \cite{forstner03, grodecka05}.
In recent experimental studies, the influence of LA phonons on the coherent QD dynamics has been measured by the tunnel charge through the QD (proportional to the excitonic population) following optical excitation \cite{ramsay10}, and also through the direct coupling of the QD's dipole to optical modes in microcavities \cite{laucht09,hohenester09}. These studies confirm that a 2LS model with a perturbative treatment of the LA phonon interaction provides an excellent description in this regime.

Previous theoretical studies of the exciton-phonon interaction in the Rabi regime have fully discarded all information about the state of the phonon bath \cite{krummheuer02,forstner03,grodecka05, vagov04,vagov07,gauger08}. Here we develop a technique for tracking the number of excitations in the environment, allowing us to analyze the net rate of absorbed or released bath energy, and we show that a continuously driven excitonic qubit constitutes a controllable two-way heat pump. Exploiting this cooling effect would help with gaining further experimental insight into the exciton-phonon coupling,  crucial for managing decoherence of semiconductor charge qubits, as well as providing an easy preparatory qubit initialization step for quantum information processing.

\begin{figure}
\includegraphics[width=0.85\columnwidth]{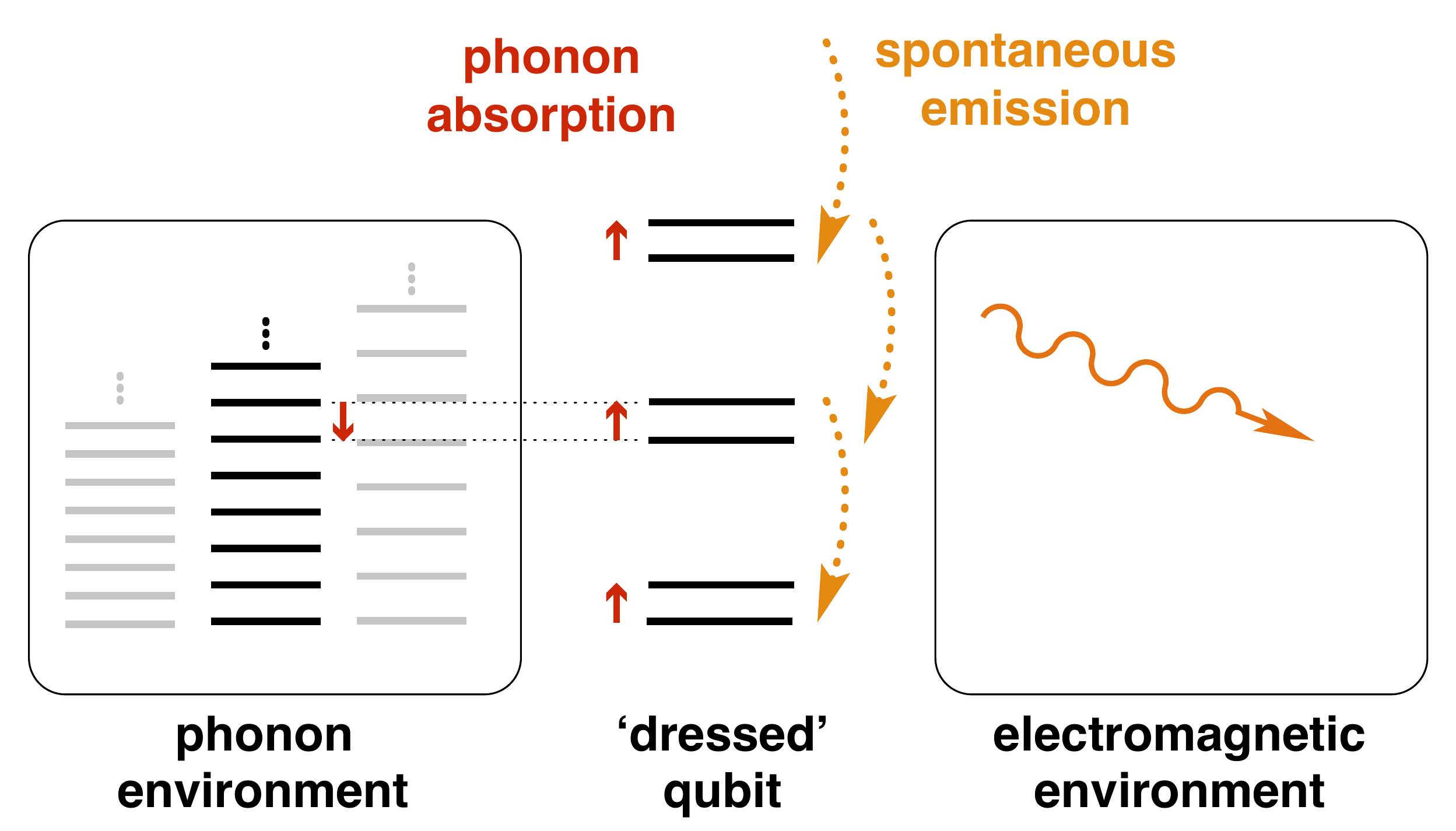} 
\caption{(color online) The driven exciton acts as a heat pump between the phonon and the electromagnetic
environment. If the driving is detuned from the exciton transition, the laser-dressed eigenstates (see Ref.~\cite{gauger08}) are composed of different amounts of the ground and excited state. Cooling is possible when spontaneous emission repeatedly takes the system into its lower eigenstate, from where it can only absorb phonons.}
\label{fig:schematic}
\end{figure}

\textit{Model} - We consider a self-assembled QD (such as InGaAs encased
in GaAs substrate) illuminated by a laser beam with frequency $\omega_{l}$,
which is nearly resonant with the crystal ground state to exciton
transition. In a frame rotating with the laser frequency and within
the rotating wave approximation (RWA), the system is governed by the
Hamiltonian ($\hbar=1$) \begin{equation}
H_{S}=\Delta/2\,\sz+\Omega/2\,\sx,\label{eq:SystemHamiltonianPM}\end{equation}
 where the detuning $\Delta$ describes the energy difference between
the basis states in the rotating frame, $\Omega$ is the Rabi frequency,
coupling the two basis states, and where $\sx=\kb{g}{e}+\kb{e}{g}$
and $\sz=\kb{e}{e}-\kb{g}{g}$ are the Pauli pseudo-spin operators
defined with respect to the QD ground state \ket{g} and single-exciton
state \ket{e}.

Let us further assume that the excitonic qubit is coupled to a bath
of phonons, resulting in the total Hamiltonian \begin{equation}
H=H_{S}+H_{B}+H_{I},\end{equation}
 where $H_{B}=\sum_{\mathbf{q}}\omega_{\mathbf{q}}\cre{a}_{\vec{q}}\an{a}_{\vec{q}}$
is the free Hamiltonian of the phonon modes with $\cre{a}_{\vec{q}}$
and $\an{a}_{\vec{q}}$ being the creation and annihilation operators
of a phonon in mode \vec{q} with frequency $\omega_{\vec{q}}$.
The exciton-phonon interaction term is generically given by \citep{mahan00,krummheuer02}
\begin{equation}
H_{I}=\sz\sum\limits _{\vec{q}}g_{\vec{q}}(\cre{a}_{\vec{q}}+\an{a}_{\vec{q}}),\label{eq:PhononInteractionPM}\end{equation}
 where the $g_{\vec{q}}$ are coupling constants. Moving into the
interaction picture with respect to $H_{S}$ and $H_{B}$ we obtain
for the transformed interaction Hamiltonian \begin{equation}
\tilde{H}_{I}(t)=\tilde{\sigma}_{z}(t)\sum\limits _{\vec{q}}g_{\vec{q}}(\cre{a}_{\vec{q}}e^{i\omega_{\vec{q}}t}+\an{a}_{\vec{q}}e^{-i\omega_{\vec{q}}t}),\label{eq:IntHamiltonianPM}\end{equation}
 where $\tilde{\sigma}_{z}(t)$ denotes the transformed $\sz$ operator.

For monitoring phonon excitations in the QD's solid state environment,
we adapt a technique from the literature on observing single electron
charge transport \citep{levitov96,shelankov03,wabnig06} by performing
a gauge transformation which adds phase markers to the interaction
Hamiltonian. Crucially, these do not alter the system's dynamics.
We define the phonon-number specific density matrix, i.e.~an object
that contains information about both the qubit dynamics and the number
of phonons in the environment, \begin{equation}
\rho_{m}=\mathrm{tr}_{E}(P_{m}\varrho),\label{eq:ProjectedRhoPM}\end{equation}
where $\mathrm{tr}_{E}$ denotes the trace over the phonon modes, $\varrho$ is the density matrix of the combined phonon-qubit
system and $P_{m}$ is a projection operator, \[
P_{m}=\frac{1}{2\pi}\int\limits _{0}^{2\pi}\mathrm{d}\lambda~e^{-im\lambda}\mathcal{E}_{\lambda},\qquad\mathcal{E}_{\lambda}=e^{i\lambda\hat{N}},\]
with the total phonon number operator $\hat{N}=\sum_{\vec{q}}\cre{a}_{\vec{q}}\an{a}_{\vec{q}}$,
and the gauge transformation operator $\mathcal{E}_{\lambda}$. The
projection operator $P_{m}$ projects out all parts of the wavefunction
with a phonon number different from $m$ (for a detailed discussion
see Ref.~\citep{shelankov03}). The probability of having emitted
$m$ phonons into the environment is then obtained by the matrix trace of the phonon-number
specific density matrix \begin{equation}
p_{m}=\mathrm{tr}(\rho_{m}).\end{equation}
 It is convenient to introduce the Fourier transformed density matrix
as\[
\rho_{\lambda}=\sum_{m}e^{im\lambda}\rho_{m}=\mathrm{tr}(\varrho_{\lambda}),\]
with the definition $\varrho_{\lambda}=\mathcal{E}_{\lambda/2}\varrho\mathcal{E}_{-\lambda/2}^{\dagger}.$
This particular choice corresponds to an initial state with a definite
phonon number as in Ref.~\citep{levitov96}. It can be shown that
the full density matrix $\varrho_{\lambda}$ in the interaction picture
obeys the von Neumann equation, \[
\dot{\varrho}_{\lambda}(t)=-i\left(\tilde{H}_{\lambda}(t)\varrho_{\lambda}(t)-\varrho_{\lambda}(t)\tilde{H}_{-\lambda}(t)\right)=\mathcal{L}_{\lambda}(t)\varrho_{\lambda}(t),\]
defining the super-operator $\mathcal{L}_{\lambda}(t)$, where the
interaction Hamiltonian $\tilde{H}_{\lambda}=\mathcal{E}_{\lambda/2}\tilde{H}_{I}\mathcal{E}_{-\lambda/2}^{\dagger}$
has acquired phase markers on the phonon creation and annihilation
operators: \begin{equation}
\tilde{H}_{\lambda}=\tilde{\sigma}_{z}(t)\sum\limits _{\vec{q}}g_{\vec{q}}(e^{-i\frac{\lambda}{2}}\cre{a}_{\vec{q}}e^{i\omega_{\vec{q}}t}+e^{i\frac{\lambda}{2}}\an{a}_{\vec{q}}e^{-i\omega_{\vec{q}}t}),\label{eq:PhononInteractionMarkersPM}\end{equation}
written in a shorter notation as $\tilde{H}_{\lambda}=\tilde{\sigma}_{z}(t)B_{\lambda}(t)$.
The phase factor $\exp(-i\lambda/2)$ then keeps track of the creation
of phonons, while $\exp(i\lambda/2)$ tracks annihilation processes.

We proceed along the standard path of deriving a master equation (ME)
\citep{breuer02}, and obtain an integro-differential equation
for the reduced density matrix of the system $\rho_{\lambda}$,\[
\dot{\rho}_{\lambda}(t)=\mathrm{tr}_{E}\int\limits _{0}^{t}ds\,\mathcal{L}_{\lambda}(t)\mathcal{L}_{\lambda}(s)\varrho_{\lambda}.\]
 After the Born-Markov approximation \cite{breuer02} the resulting
Markovian ME reads\[
\dot{\rho}_{\lambda}(t)=-\int_{0}^{\infty}ds\,\sum_{i,j}G_{ij}^{\lambda}(s)\mathcal{S}_{i}(t)\mathcal{S}_{j}(t-s)\rho_{\lambda}(t),\]
where we have defined the super operators $\mathcal{S}_{1}(t)X=\tilde{\sigma}_{z}(t)X$,
$\mathcal{S}_{2}(t)X=X\tilde{\sigma}_{z}(t)$ and the environment
correlation functions $G_{11}^{\lambda}(s)=\exx{B_{\lambda}(s)B_{\lambda}(0)}$,
$G_{12}^{\lambda}(s)=-\exx{B_{-\lambda}(0)B_{\lambda}(s)}$, $G_{21}^{\lambda}(s)=-\exx{B_{-\lambda}(s)B_{\lambda}(0)}$
and $G_{22}^{\lambda}(s)=\exx{B_{-\lambda}(0)B_{-\lambda}(s)}$. To
simplify the further algebraic evaluation, we express the operator
$\tilde{\sigma}_{z}(t)$ in terms of system eigenoperators, yielding
\begin{equation}
\tilde{\sigma}_{z}(t)=\sum_{\omega\in\{0,\Lambda\}}\left(e^{-i\omega t}P_{\omega}+e^{i\omega t}P_{\omega}^{\dagger}\right),\label{eq:InstSystemOpsPM}\end{equation}
where $\Lambda=\sqrt{\Omega^{2}+\Delta^{2}}$ now denotes the spacing
between the system eigenstates $\ket{-}=\cos\theta\ket{g}-\sin\theta\ket{e}$
and $\ket{+}=\cos\theta\ket{e}+\sin\theta\ket{g}$ of Hamiltonian
(\ref{eq:SystemHamiltonianPM}) with $2\theta=\arctan\Omega/\Delta$.
In this basis, $P_{0}=\cos2\theta(\kb{-}{-}-\kb{+}{+})/2$ and $P_{\Lambda}=\sin2\theta\kb{-}{+}$.

By introducing the Hermitian operator $\tilde{\mathcal{P}}=2P_{0}+e^{-i\Lambda t}P_{\Lambda}+e^{i\Lambda t}P_{\Lambda}^{\dagger}$
and abbreviating $\tilde{\mathcal{Q}}=e^{-i\Lambda t}P_{\Lambda}$,
we obtain the following interaction picture ME after some straightforward
algebra \begin{eqnarray*}
\dot{\rho}_{\lambda} & = & \Gamma_{\downarrow}\left(e^{-i\lambda}\left(\tilde{\mathcal{Q}}\rho_{\lambda}\tilde{\mathcal{P}}^{\dagger}+\tilde{\mathcal{P}}\rho_{\lambda}\tilde{\mathcal{Q}}^{\dagger}\right)-\tilde{\mathcal{P}}^{\dagger}\tilde{\mathcal{Q}}\rho_{\lambda}-\rho_{\lambda}\tilde{\mathcal{Q}}^{\dagger}\tilde{\mathcal{P}}\right)\\
 & + & \Gamma_{\uparrow}\left(e^{i\lambda}\left(\tilde{\mathcal{P}}^{\dagger}\rho_{\lambda}\tilde{\mathcal{Q}}+\tilde{\mathcal{Q}}^{\dagger}\rho_{\lambda}\tilde{\mathcal{P}}\right)-\tilde{\mathcal{P}}\tilde{\mathcal{Q}}^{\dagger}\rho_{\lambda}-\rho_{\lambda}\tilde{\mathcal{Q}}\tilde{\mathcal{P}}^{\dagger}\right),\end{eqnarray*}
where the rates are given by $\Gamma_{\downarrow}=J(\Lambda)\left(n(\Lambda)+1\right)/2$
and $\Gamma_{\uparrow}=J(\Lambda)n(\Lambda)/2$ when the phonon bath
$\rho_{E}$ is in a thermal state %
\footnote{A small renormalization due to principal value terms in the ME can
be absorbed in the free evolution of the qubit.%
}. Further, \begin{equation}
J(\omega)=2\pi\sum_{\vec{q}}\abs{g_{\vec{q}}}^{2}\delta(\omega-\omega_{\vec{q}})\end{equation}
is the spectral density of phonon modes and $n(\omega)=\left(\exp(\beta\omega)-1\right)^{-1}$
denotes the thermal occupancy of a phonon mode with frequency $\omega$.

Upon going back to the Schr\"odinger picture, the operators $\tilde{\mathcal{P}}$
and $\tilde{\mathcal{Q}}$ lose their time-dependent phase factors,
turning into $\mathcal{P}=\sz$ and $\mathcal{Q}=P_{\Lambda}$, respectively.
We can return to the number representation with the transformation
\[
\rho_{m}=\frac{1}{2\pi}\int_{0}^{2\pi}\mathrm{d}\lambda~e^{-im\lambda}\rho_{\lambda}.\]
 This yields set of coupled differential equations for the evolution
of phonon-specific density matrices. After a RWA (implemented by setting
$\mathcal{P}=\mathcal{L}$, expected to be valid whenever $\Lambda\gg J(\Lambda)$
\citep{breuer02,stace05}), we finally obtain a ME in diagonal Lindblad
form \begin{eqnarray}
\dot{\rho}_{m} & = & -i[H_{S}(t),\rho_{m}]\nonumber \\
 & + & \Gamma_{\downarrow}\left(2P_{\Lambda}\rho_{m+1}P_{\Lambda}^{\dagger}-P_{\Lambda}^{\dagger}P_{\Lambda}\rho_{m}-\rho_{m}P_{\Lambda}^{\dagger}P_{\Lambda}\right)\nonumber \\
 & + & \Gamma_{\uparrow}\left(2P_{\Lambda}^{\dagger}\rho_{m-1}P_{\Lambda}-P_{\Lambda}P_{\Lambda}^{\dagger}\rho_{m}-\rho_{m}P_{\Lambda}P_{\Lambda}^{\dagger}\right).~~\label{eq:DiagNumberMEPM}\end{eqnarray}
This equation describes the joint phonon-qubit dynamics in the case
where the phonon environment is only weakly disturbed from thermal
equilibrium by the excitonic qubit.

\textit{Phonon-assisted transitions} - We now apply Eq.~(\ref{eq:DiagNumberMEPM})
to an excitonic qubit that is optically driven by a pulse of constant
intensity $\Omega$. The initial state at $t=0$ is the system ground
state with zero excitations in the environment: $\rho_{n}(0)=\delta_{n,0}\kb{g}{g}$.
For simplicity, we define the spectral density phenomenologically:
$J(\omega)=\alpha\omega^{3}\exp\left(-\omega^{2}/\omega_{c}^{2}\right)$,
where $\alpha$ describes the effective electron-phonon coupling strength
and $\omega_{c}$ is the high frequency phonon cut-off. For relatively
weak driving with a peak Rabi frequency $\Lambda$ well below both
the electron and the hole cut-off, we can neglect the exponential
cut-off term altogether. Setting $\alpha=1/4~\mathrm{ps}^{2}$ yields
a coupling strength that is consistent with the magnitude of the GaAs
deformation potential reported in the literature \citep{takagahara99,krummheuer02}.

The structure of Eq. (\ref{eq:DiagNumberMEPM}) permits the emission
or absorption of no more than a single phonon: The system is initialised
in the state $\kb{g}{g}=(\ket{-}+\ket{+})(\bra{-}+\bra{+})/2$, i.e.
in an equal superposition of system eigenstates. The Lindblad operator
$P_{\Lambda}$ induces a transition from \ket{+} to \ket{-} while
$P_{\Lambda}^{\dagger}$ raises population from \ket{-} to \ket{+}.
Once a decay process has happened, we find the system in the \ket{-}
state, meaning it cannot decay again.

Provided the excitation is sufficiently long, the population ratios
thus tend to a Boltzmann distribution, as is obvious from the phonon
emission rate proportional to $n(\Lambda)+1$ and the absorption rate
proportional to $n(\Lambda)$, \[
\lim_{t\to\infty}\frac{\mathrm{tr}(\rho_{0}(t)\kb{+}{+})}{\mathrm{tr}(\rho_{1}(t))}=\lim_{t\to\infty}\frac{\mathrm{tr}(\rho_{-1}(t))}{\mathrm{tr}(\rho_{0}(t)\kb{-}{-})}=e^{-\beta\Lambda}.\]

\begin{figure}
\begin{centering}
\includegraphics[width=1\linewidth]{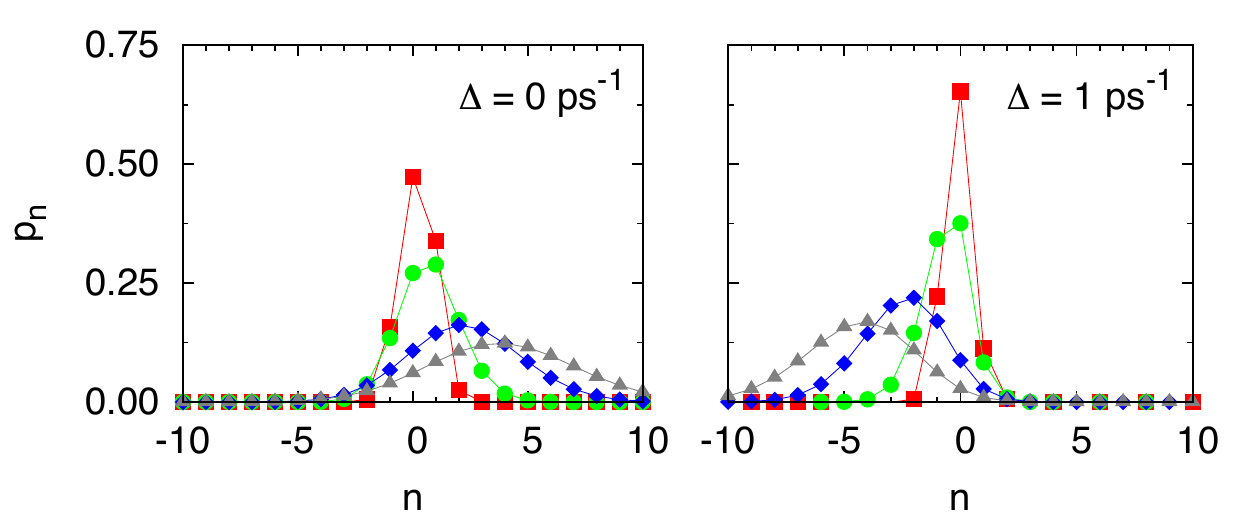} 
\par\end{centering}

\caption{(color online) The distribution of $p_{n}$ at different points of time in the presence
of radiative decay. The constant excitation pulse uses $\Omega=1~\mathrm{ps}^{-1}~(\approx2/3~\mathrm{meV})$
and a detuning $\Delta$ as indicated in the plots (in $\mathrm{ps}^{-1}$).
The decay rate is fixed at $\Gamma=0.1~\mathrm{ps}^{-1}$. The different
colors/symbols correspond to time as follows: red squares = 1.2, green
circles = 10, blue diamonds = 40, gray triangles = 70, all in units of
full Rabi cycles, $2\pi/\sqrt{\Omega^{2}+\Delta^{2}}$.}

\label{fig:DistributionPM} 
\end{figure}

So far, the only perturbation to the system has been caused by the
coupling to the phonon bath, resulting in single phonon emission and
absorption. Realistically, other dissipative processes will be present
in any physical systems. We shall include these using additional phenomenological
noise operators, such as pure dephasing and radiative decay of the
exciton. We model these processes with an additional Lindblad dissipator
\citep{breuer02} on the right-hand side of Eq.~(\ref{eq:DiagNumberMEPM}),
\begin{equation} 
\mathcal{D}\rho=\Gamma\left(L\rho L^{\dagger}-\frac{1}{2}(L^{\dagger}L\rho+\rho L^{\dagger}L)\right),\label{eq:ExtraLindbladiansPM}
\end{equation}
with respective Lindblad operators $L=\sz$ and $L=\sm$, and where $\Gamma$ is the dephasing or decay rate. These operators
do not preserve the system's eigenstates; consequently, under their action, phonon-assisted transitions
become possible in any of the $\rho_{n}$ subspaces. This leads to
a dynamic equilibrium, where phonon-assisted transitions keep occurring
after the transient (coherent) evolution of $\rho=\sum_{n}\rho_{n}$
has subsided. Our theoretical model is well-suited for illustrating
this behaviour: Fig. \ref{fig:DistributionPM} presents the $p_{n}$
distribution at different points of time for the case of optical decay.

The pure dephasing $\sz$ operator randomises the phase between \ket{g}
and \ket{e}, thus balancing the population of \ket{-} and \ket{+}.
Consequently, once the steady state has been reached, phonon emission
always occurs with a faster rate then absorption, and the distribution
is therefore shifted in the direction of increasing $n$, meaning
the average number of emitted phonons increases. On the other hand,
the distribution can move in either direction for optical decay
from \ket{e} to \ket{g}. For $\Delta=0$, \ket{g} consists of
an equal superposition of \ket{-} and \ket{+}, while it contains
a larger \ket{-} component for $\abs{\Delta}>0$. Under these latter
circumstances, it is possible for phonon absorption to permanently
dominate over emission, shifting the distribution in the direction
of decreasing $n$, as shown in the right panel of Fig. \ref{fig:DistributionPM}.
In this case, thermal energy is removed from the QD's bulk surroundings
and released into the wider environment by spontaneous photon emission (depicted in Fig.~\ref{fig:schematic}).

\textit{Heat transfer rate} - To quantify this observation, we consider only the radiative decay operator of Eq.~(\ref{eq:ExtraLindbladiansPM})
in the following. We proceed by calculating the rate of phonon emission
or absorption, which is given by \citep{wabnig06}: \begin{eqnarray}
\exx{\dot{n}(t)} & = & \ddt\sum_{m}m\: p_{m}=\left.i\frac{\mathrm{d}}{\mathrm{d}\lambda}\mathrm{tr}(\dot{\rho}_{\lambda})\right|_{\lambda=0}\nonumber \\
 & = & 2~\mathrm{tr}\left(\Gamma_{\downarrow}\mathcal{Q}\rho(t)\mathcal{Q}^{\dagger}-\Gamma_{\uparrow}\mathcal{Q}^{\dagger}\rho(t)\mathcal{Q}\right),\label{eq:PhononRatePM}\end{eqnarray}
where $\rho(t)$ is obtained by integrating Eq. (\ref{eq:DiagNumberMEPM})
inclusive of the Lindblad dissipator (\ref{eq:ExtraLindbladiansPM})
with $L=\sm$ and disregarding the indices $m$ of $\rho_{m}$. %
\begin{figure}
\begin{centering}
\includegraphics[width=1\linewidth]{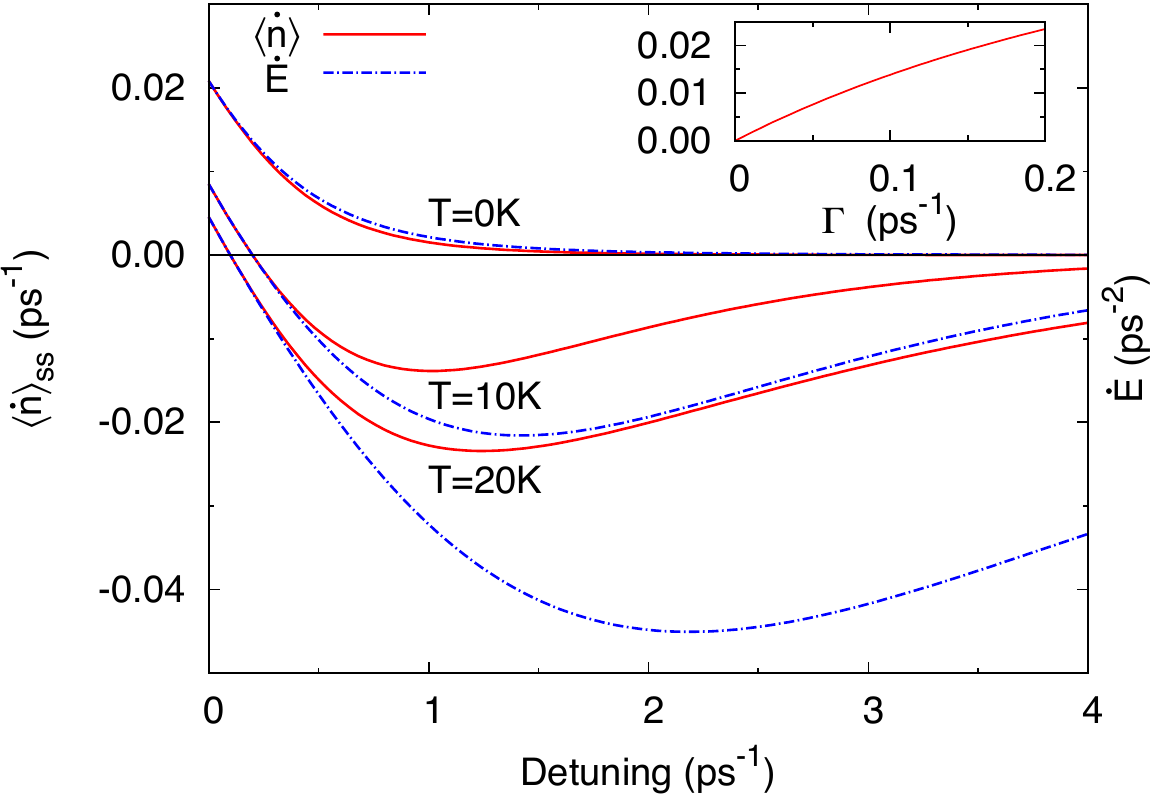} 
\par\end{centering}

\caption{(color online) The rate of phonon-induced transitions $\langle\dot{n}\rangle_{ss}$
(red) and the rate of energy transfer $\dot{E}$ (blue), both as a
function of the detuning for $\Omega=1~\mathrm{ps}^{-1}$. The rates
shown here are the steady state values of Eq. (\ref{eq:PhononRatePM})
for the decay rate $\Gamma=0.1~\mathrm{ps}^{-1}$. A positive sign
indicates net phonon emission whereas a negative corresponds to net
phonon absorption. The inset shows $\exx{\dot{n}}_{ss}$ for fixed
$\Delta=1~\mathrm{ps}^{-1}$ as a function of $\Gamma$ at $T=10~\mathrm{K}$.}

\label{fig:CoolingRatePM} 
\end{figure}

Fig. \ref{fig:CoolingRatePM} presents the steady-state value $\exx{\dot{n}}_{ss}$
of Eq. (\ref{eq:PhononRatePM}) as a function of $\Delta$. As expected,
net phonon absorption is only possible at finite temperature for off-resonant
excitation. It is a natural question to ask at which rate energy is
transferred to or from the surroundings of the system. Considering
the quantity $\dot{E}=\sqrt{\Omega^{2}+\Delta^{2}}\;\exx{\dot{n}}_{ss}$
shows that the number of absorbed phonons decreases for a larger detuning,
yet their energy is greater, shifting the $\Delta$ that achieves
optimal heat transfer. The inset of Fig. \ref{fig:CoolingRatePM}
shows that the process is limited by the radiative decay time $\Gamma$
of the system. A saturation only occurs when the spontaneous emission
rate becomes faster than that of phonon-mediated transitions, but this
regime would require an unrealistic optical lifetime of the order
of a picosecond or less. 

We proceed by estimating the achievable cooling rate for realistic
parameters in SI units. Using $\Omega_{0}=\Delta=1~\mathrm{ps}^{-1}$
corresponds to a phonon energy of $\Lambda=1.49\times10^{-22}~\mathrm{J}$.
With approximately 0.02 absorbed phonons per picosecond (at $T=20~\mathrm{K}$),
we obtain a theoretical energy transfer rate of roughly $3\times10^{-12}~\mathrm{J/s}$.
Neglecting heating effects, this achieves a temperature reduction
of the order of one Kelvin per second for a micrometer cube of GaAs
\footnote{Using a mass density of $\mu=5.3\times10^{3}~\mathrm{kg/m}^{3}$ and
a specific heat of $350~\mathrm{J/(kg\; K)}$ gives a heat
capacity of $1.85\times10^{-12}~\mathrm{J/K}$ for a micrometer cube of GaAs.}.
Of course, any real sample will also be subject to heating processed,
e.g.~by the laser illumination and by thermal contact to its surroundings.
While it is difficult to estimate the precise rate of heating, it
could plausibly exceed the cooling rate in bulk samples. However,
by incorporating the QD on a nanopillar or a cantilever, the thermal
coupling to the bulk could be reduced to make net cooling possible
\footnote{This may also require a straightforward substitution of $J(\omega)$
to account for a modified phonon spectrum.%
}. An experiment to show the cooling effect could proceed in the following
way: insert a cooling cycle before the standard sequence for observing
Rabi oscillations and study the influence of the length of the cooling
cycle on coherence times. If cooling is successful one would expect
increased coherence times.

\textit{Summary} - We have shown that a single excitonic QD, an experimentally
well studied system, can act as a heat pump. Adapting a counting statistic
approach from the context of charge transfer, 
 we have discussed how energy can be removed from the QD's environment
by means of repeated phonon absorption in conjunction with spontaneous
photon emission. This effect does not rely on the structure of the
spectral density, making our analysis applicable to similar systems
with a different coupling mechanism to a bath of harmonic oscillators.
This opens up the possibility of experimental investigation of quantum
heat pumps as well as the prospect of environment preparation for the
use of excitonic systems in quantum information processing.

\textit{Acknowledgements} - We thank Mete Atature and Brendon Lovett
for stimulating discussions. This work was supported by the Marie
Curie Early Stage Training network QIPEST (MESTCT-2005-020505) and
the QIP IRC (GR/S82176/01). JW thanks The Wenner-Gren Foundations
for financial support.


\end{document}